\documentclass[aps,pra,superscriptaddress,twocolumn,floatfix,showpacs,a4paper]{revtex4}

\usepackage{graphicx,graphics,epsfig}   
\usepackage{dcolumn}    
\usepackage{bm}         
\usepackage{amsmath}    
\usepackage{verbatim}   
\usepackage{color}      
\usepackage{subfigure}  
\usepackage{times,natbib}
\usepackage{amsmath,amsfonts,amssymb,graphics,graphicx,epsfig,color,times,natbib}

\usepackage{amssymb}
\usepackage{mathrsfs}
\usepackage{amsfonts}
\usepackage{amsmath,graphicx,epsfig,color,CJK}
\usepackage{pifont,bbding,amssymb}
\usepackage[hypertex]{hyperref}

\begin{document}
\title{Detecting Full $N$-Particle Entanglement in Arbitrarily High-Dimensional Systems with Bell-Type Inequality}

\author{Jing-Ling Chen}
 \email{chenjl@nankai.edu.cn}
\affiliation{Theoretical Physics Division, Chern Institute of
Mathematics, Nankai University, Tianjin 300071, People's Republic of
China}

\author{Dong-Ling Deng}
\affiliation{Theoretical Physics Division, Chern Institute of
Mathematics, Nankai University, Tianjin 300071, People's Republic of
China} \affiliation{Department of Physics and MCTP, University of
Michigan, Ann Arbor, Michigan 48109, USA}

\author{Hong-Yi Su}
\affiliation{Theoretical Physics Division, Chern Institute of
Mathematics, Nankai University, Tianjin 300071, People's Republic of
China}

\author{Chunfeng Wu}
\affiliation{Department of Physics and Centre
for Quantum Technologies, National
University of Singapore, 2 Science Drive 3, Singapore 117542}

\author{C. H. Oh}
 \email{phyohch@nus.edu.sg}
\affiliation{Department of Physics and Centre for Quantum
Technologies, National University of Singapore, 2 Science Drive 3,
Singapore 117542}

\date{\today}

\begin{abstract}
We derive a set of Bell-type inequalities for arbitrarily
high-dimensional systems, based on the assumption of partial
separability in the hybrid local-nonlocal hidden variable model.
Partially entangled states would not violate the inequalities, and
thus upon violation, these Bell-type inequalities are sufficient
conditions to detect the full $N$-particle entanglement and invalidity
of the hybrid local-nonlocal hidden variable description.
\end{abstract}



\maketitle

\section{introduction}
Entanglement is one of the most fundamental features of quantum
mechanics, and it lies at the
heart of recent quantum information theory. As a result, many
remarkable achievements, such as quantum teleportation
\cite{Benne1} and the higher levels of security in cryptography
\cite{Gisin1} have been attained owing to the quantum
entanglement.

Given $N$-particle quantum systems, the correlations among them have
been the subject of several recent studies
\cite{1987Svetlichny,2002Seevinck,2002Collins,2009Ghose,Rpreprint,2005M.
Zukowski}. This is also motivated by the question whether the
correlations in recent experiments on 3- or 4-particle systems are
due to the full $N$-particle entanglement and not just combinations
of quantum entanglement of smaller number of particles
\cite{Rauschenbeutel,JWPan}. For $N=2$, the entanglement type of the
bipartite system is humdrum, i.e., it is either entangled or
separable. However, the situation is dramatically changed when $N\ge
3$, besides the totally separable states, there are partially
entangled states and fully $N$-particle entangled states. Consider
all possible decompositions of a $N$-particle state as a mixture of
pure states $\rho^{N}=\sum_i p_i|\Psi_i\rangle\langle\Psi_i|$, if
for any decomposition there is at least one $|\Psi_i\rangle$ showing
$N$-particle entanglement, then we shall say that $\rho^{N}$
exhibits full $N$-particle entanglement; if the state is not
separable or not fully $N$-particle
entangled, then we call that $\rho^{N}$ is partially entangled. %
%
%
%
%

Entanglement has been studied extensively in connection with the
Bell inequality. The Bell inequality was originally proposed to
ruled out local realism description of quantum
mechanics~\cite{Bell}. This presents us a concept, the so-called
nonlocality, which is revealed by violations of the Bell inequality.
Generally, entanglement and nonlocality are two different concepts.
Although there exist entangled states not violating the known Bell
inequalities, the violation of Bell inequality means the studied
system is entangled, allowing us to detect entanglement. The
conventional ``$N$-particle Bell inequalities" are designed to deny
the local hidden variable (LHV) models
\cite{MABK,Werner1,Zukowski1}. For $N$-particle quantum systems, the
partially entangled states and the fully $N$-particle entangled
states may violate the same Bell inequality, namely, the
conventional Bell inequalities do not distinguish at all the
partially entangled states and the fully $N$-particle entangled
states. Actually, a particle can decay into several particles, this
gives rise to a natural question: Are the resultant decaying systems
in a fully entangled state or just a partially entangled state? In
1987, Svetlichny triggered the problem and proposed a Bell-type
inequality to distinguish full three-qubit entanglement from
partially two-qubit ones~\cite{1987Svetlichny}. The Svetlichny
inequality is essentially different from the conventional Bell
inequality, because the former is designed for a hybrid
local-nonlocal hidden variable (HLNHV) model and the latter is for a
LHV model. As the name implies, HLNHV models utilize the fusion of
local and nonlocal descriptions based on the assumption of partial
separability. Fifteen years later, Seevinck \emph{et al.} and
Collins \emph{et al.} independently generalized the Svetlichny
inequality from three-qubit case to arbitrarily $N$-qubit case
~\cite{2002Seevinck,2002Collins}. Upon violation, these $N$-qubit
Bell-type inequalities are sufficient conditions for detecting full
$N$-qubit entanglement.

In this paper, we generalize the Svetlichny inequalities from the
qubit case to $N$ arbitrarily $d$-dimensional systems ($N$-qudit).
These Bell-type inequalities are derived based on the assumption of
partial separability, or more generally speaking, on the so-called
HLNHV models, thus the quantum mechanical violations of these
inequalities provide experimentally accessible conditions to detect
the full $N$-qudit entanglement and rule out the HLNHV models. The
paper is organized as follows. We present the $N$-qudit Bell type
inequality based on the HLNHV model in Sec. \ref{s2}. The proof of
the $N$-qudit inequality begins with the cases of $N=3,4$, and the
result is generalized to the case of arbitrary $N$. In Sec.
\ref{s3}, we investigate quantum violation of the $N$-qudit
inequality. We show that the Greenberger-Horne-Zeilinger (GHZ)
states violate our Bell type inequality and find the explicit form
of the violation which depends on particle number $N$ and dimension
$d$. We also investigate noise resistance of the inequality by using
the so-called critical visibility. It is found that our Bell-type
inequality is more noise resistant than the Svetlichny one for $N$
qudits when $d\geq 3$. We end with conclusions in the last section.

\section{$N$-qudit Bell-type inequality}
\label{s2}
Consider an experimental situation involving $N$ particles in which
two measurements $i_n=1,2 \;(n=1,\cdots,N$) can be performed on
each particle. Each of the measurements has $d$ possible outcomes:
$x_{i_n}=0,1,\cdots,d-1$. We now follow Svetlichny's
splendid ideas~\cite{2002Seevinck,1987Svetlichny} and make the
following assumption of partial separability: The $N$-qudit system
is composed of many subsystems, which might be correlated in any way
(e.g. entangled) but are uncorrelated with respect to each other.
Since we can always take any two subsystems jointly as a single one but
still uncorrelated with respect to the others, we only need to focus
on the case that the composed system consists of only
two uncorrelated subsystems involving $m<N$ and $N-m<N$ qudits,
respectively. For simplicity, we also assume that the first
subsystem is formed by the first $m$ qudits and the other by the
remaining qudits. Denote the probability of observing the results
$x_{i_n}$ by $P(x_{i_1},x_{i_2},\cdots,x_{i_N})$, then the partial
separability assumption can be expressed as
\begin{eqnarray}
&&P(x_{i_1},x_{i_2}, \cdots, x_{i_N}) =
\int_{\Gamma}P_1(x_{i_1},x_{i_2},\cdots, x_{i_m}|\lambda)  \nonumber\\
&& \;\;\;\;\;\;\;\;\;\;\;\;\;\;\;\times
P_2(x_{i_{m+1}},x_{i_{m+2}},\cdots, x_{i_N}|\lambda)\;
\rho(\lambda)d\lambda,
\end{eqnarray}
where $P_1(x_{i_1},x_{i_2},\cdots,x_{i_m}|\lambda)$ and
$P_2(x_{i_{m+1}},\cdots,x_{i_N}|\lambda)$ are probabilities
conditioned to the hidden variable $\lambda$; $\Gamma$ is the total
$\lambda$ space and $\rho(\lambda)$ is a statistical distribution of
$\lambda$, which satisfies $\rho(\lambda)\geq0$ and $\int_\Gamma
d\lambda\rho(\lambda)=1$.  Other decompositions can be described
with a different value of $m$ and different choices of the composing
qudits. A HLNHV model can then be well defined based on the
assumption of partial separability and the formula of the
factorizable probability, readers who are interested in it may refer
to Refs.~\cite{2004Mitchell-three-particle-nonlocality,2002Collins}.
If the probability factorization can be
\begin{eqnarray}
P(x_{i_1},\cdots,x_{i_N})= \int_{\Gamma}d\lambda\rho(\lambda)
P_1(x_{i_1}|\lambda) \cdots P_N(x_{i_N}|\lambda),
\end{eqnarray}
the HLNHV model
then reduces to the usual LHV model.


For convenience, we introduce two functions:
\begin{eqnarray}\label{gfunc}
g_1(x+s_t)&=&\frac{S-M(x+s_t,d)}{S},\nonumber\\
g_2(x+s_t)&=&\frac{S-M(-x-s_t,d)}{S}.
\end{eqnarray}
Here $S=\frac{d-1}{2}$ is the spin value of the particle; $s_t$
means a shift of the argument $x$; $M(x,d)=(x,\texttt{mod}\; d)$ and
$0\leq M(x,d)\leq d-1$.
The $N$-qudit Bell-type inequality reads
\begin{eqnarray}
I^{N}=-\biggr(\sum_{i_1,i_2,\cdots,i_N=1}^2Q_{i_1i_2\cdots
i_N}\biggr)\leq2^{N-1},\label{N-qudit}
\end{eqnarray}
with
\begin{eqnarray}
Q_{i_1i_2\cdots i_N}{\equiv}\sum_{x_{i_1},
\cdots,x_{i_N}=0}^{d-1}f^{i_1i_2\cdots i_N} P(x_{i_1},
x_{i_2},\cdots,x_{i_N}).\label{correlation}
\end{eqnarray}
Let $\mathcal {I}\; {\equiv}i_1i_2{\cdots}i_N$, and $t(\mathcal
{I})$ denotes the times that the index ``2" appears in the string
$\mathcal {I}$, we then abbreviate the coefficient as follows:
\begin{eqnarray}
&&f^{\mathcal {I}}(x_{i_1}, x_{i_2},\cdots,x_{i_N},s_t)\equiv g_1\nonumber\\
&&=1-M(x_{i_1}+ \cdots+x_{i_N}+s_t,d)/S
\end{eqnarray}
if $t(\mathcal {I})$ is even and
\begin{eqnarray}
&&f^{\mathcal {I}}(x_{i_1}, x_{i_2},\cdots,x_{i_N},s_t)\equiv g_2\nonumber \\
&&=1-M(-x_{i_1}-\cdots-x_{i_N}-s_t,d)/S
\end{eqnarray}
if $t(\mathcal {I})$ is odd, and $s_t{\equiv}s_{t(\mathcal
{I})}=3\times(1-[\frac{t(\mathcal {I})}{2}])$, where $[\frac{t}{2}]$
means the integer part of $\frac{t}{2}$. The above inequality is
symmetric under permutations of the $N$ particles. Essentially, the
inequality (\ref{N-qudit}) is a kind of probabilistic Bell-type
inequality if substitute Eq. (\ref{correlation}) into inequality
(\ref{N-qudit}). We express it in the form of inequality
(\ref{N-qudit}) for two reasons: (i) to make the inequality succinct
and (ii) $Q_{\mathcal {I}}$ may be regarded as generalized
correlation functions of $N$-qudit in comparison to the typical form
of correlation functions of qubits. As for the coefficient
$f^{\mathcal {I}}$, its possible values are equal to $S_z/S \in
\{-1, -1+1/S,\cdots,1\}$, where $S_z$ is expectation value of the
$z$-component of the spin operators. Especially $f^{\mathcal {I}}$
has only two possible values $\pm 1$ when $S=1/2$, $Q_{\mathcal
{I}}$ reduces to the typical form of correlation functions of
qubits.

In the following, we shall prove that the upper bound of the
inequality is $2^{N-1}$.



\emph{Three qudits.}---Our inequality for three qudits reads
\begin{eqnarray}
I^{3}&=&-Q_{111}-Q_{112}-Q_{121}-Q_{122}\nonumber\\
&&-Q_{211}-Q_{212}-Q_{221}-Q_{222}\leq4,
\label{3-qudit}
\end{eqnarray}
with $f^{111}=f^{122}=f^{221}=f^{212}=g_1$ and
$f^{222}=f^{112}=f^{121}=f^{211}=g_2$; the shift
$s_{t(111)}=s_{t(112)}=s_{t(121)}=s_{t(211)}=3$ and
$s_{t(122)}=s_{t(212)}=s_{t(221)}=s_{t(222)}=0$.
We assume that, for three qudits, the two uncorrelated subsystems are the first two qudits
and the third qudit. Hence, in a HLNHV model with two-setting
scenario, four possible outcomes for the first two qudits
$x_{i_1}+x_{i_2}$ and two outcomes for the third qudit
$x_{i_3}$ are independent of each other. We then simply denote
$x_{i_1}+x_{i_2}$ by a single variable $\xi_{ij}$ and $x_{i_3}$ by
$\zeta_k$, both are from $0$ to $d-1$. Moreover, since any
nondeterministic local variable model can be made deterministic by
adding additional variables~\cite{1998Percival}, we only need to
consider the \emph{deterministic} versions~\cite{1982Fine} of the
HLNHV model in which for each value of $\lambda$, the measurement outcomes are completely determined, namely, the probability of
obtaining each possible outcome is either $0$ or $1$.
For each $\lambda$, we have
predetermined values for the outcomes of $\xi_{ij}$ and $\zeta_k$
$(i, j, k=1, 2)$.

Another preliminary knowledge is about our two-qudit
inequality
\begin{eqnarray}
I^{2}=-Q_{11}-Q_{12}-Q_{21}-Q_{22}\leq2,\label{originalCGLMP}
\end{eqnarray}
where $f^{11}=f^{22}=g_1, f^{12}=f^{21}=g_2$,
$s_{t(11)}=s_{t(12)}=s_{t(21)}=3$, and $s_{t(22)}=0$. From the
definition of $Q_{i_1i_2}$, we have the explicit form of inequality
(\ref{originalCGLMP}) as
\begin{eqnarray}
I^{2}=-g_1(r_{11}+3)-g_2(r_{12}+3)-g_2(r_{21}+3)-g_1(r_{22}),\label{originalCGLMP2}
\end{eqnarray}
here $r_{ij}{\equiv}\alpha_i+\beta_j$, $\alpha_i$ being the outcome
of the first qudit for the $i$-th measurement and $\beta_j$ being
that of the second qudit for the $j$-th measurement. In the
following, we show that inequality (\ref{originalCGLMP}) is an
equivalent form of the well-known
Collins-Gisin-Linden-Massar-Popescu (CGLMP) inequality
\cite{2002CGLMPBI}, which is usually of the form
\begin{eqnarray}
I_{\rm CGLMP}=Q_{11}+Q_{12}+Q_{21}-Q_{22}\leq2.
\end{eqnarray}
In our notation, 
the CGLMP inequality can be rewritten as
\begin{eqnarray}
I_{\rm CGLMP}=g_2(r'_{11})+g_1(r'_{12})+g_1(r'_{21})-g_1(r'_{22}),
\end{eqnarray}
with $r'_{ij}{\equiv}\alpha'_i+\beta'_j$. By using
$g_1(x)=-g_2(x+1)$, inequality (\ref{originalCGLMP}) [or (\ref{originalCGLMP2})] is of the form
\begin{eqnarray}
I^{2}=g_2(r_{11}+4)+g_1(r_{12}+2)+g_1(r_{21}+2)-g_1(r_{22}).
\end{eqnarray}
Let $\alpha'_1=\alpha_1+2$, $\alpha'_2=\alpha_2$, $\beta'_1=\beta_1+2$,
and $\beta'_2=\beta_2$, one immediately finds that $I^{2}$ and
$I_{\rm CGLMP}$ are of the same form. Hereafter we simply
call~inequality (\ref{originalCGLMP}) as the CGLMP inequality. The
proof of inequality (\ref{N-qudit}) resorts to $I^{2}\leq 2$.

To prove $I^{3} \le 4$, we write $I^3=\mathbb{I}_1+\mathbb{I}_2$ with
\begin{eqnarray}
&&\mathbb{I}_1{\equiv}-Q_{111}-Q_{112}-Q_{121}-Q_{122}\nonumber\\
&&\mathbb{I}_2{\equiv}-Q_{211}-Q_{212}-Q_{221}-Q_{222}.
\end{eqnarray}
According to the definition of $Q_{i_1i_2i_3}$, we have
\begin{eqnarray}
\mathbb{I}_1&=&-g_1(\xi_{11}+\zeta_1+3)-g_2(\xi_{11}+\zeta_2+3)\nonumber\\
&&-g_2(\xi_{12}+\zeta_1+3)-g_1(\xi_{12}+\zeta_2).
\end{eqnarray}
From Eq. (\ref{originalCGLMP2}), if one sets $\xi_{11}=\alpha_1$, $\xi_{12}=\alpha_2$,
$\zeta_1=\beta_1$, $\zeta_2=\beta_2$, then one easily finds that
$\mathbb{I}_1$ is equivalent to $I^{2}$, thus $\mathbb{I}_1\leq2$.
Similarly, we find
\begin{eqnarray}
\mathbb{I}_2&=&-g_2(\xi_{21}+\zeta_1+3)-g_1(\xi_{21}+\zeta_2)\nonumber \\
&&-g_1(\xi_{22}+\zeta_1)-g_2(\xi_{22}+\zeta_2),
\end{eqnarray} 
and set $\xi_{22}=\alpha_1+3$, $\xi_{21}=\alpha_2$,
$\zeta_1=\beta_1$, $\zeta_2=\beta_2$, then $\mathbb{I}_2$ is
equivalent to $I^{2}$, so $\mathbb{I}_2\leq2$. Thus we have
$I^{3}=\mathbb{I}_1+\mathbb{I}_2\leq{4}$.


\emph{Four qudits.}---Our inequality for four qudits reads
\begin{eqnarray}
I^{4}&=&-Q_{1111}-Q_{1112}-Q_{1121}-Q_{1211}-Q_{2111}\nonumber\\
&&-Q_{1122}-Q_{2211}-Q_{1212}-Q_{2121}-Q_{1221}\nonumber\\
&&-Q_{2112}-Q_{1222}-Q_{2122}-Q_{2212}-Q_{2221}\nonumber\\
&&-Q_{2222}\leq8,
\end{eqnarray}
with
$f^{1111}=f^{1122}=f^{2211}=f^{1212}=f^{2121}=f^{1221}=f^{2112}=f^{2222}=g_1$, and the
others are $g_2$;
the shift
$s_{t(1111)}=s_{t(1112)}=s_{t(1121)}=s_{t(1211)}=s_{t(2111)}=3$,
$s_{t(2222)}=-3$, and the others are zero. The process to obtain the
upper bound is similar to that of the three qudits. We first write the 4-qudit inequality as
$I^4=\mathbb{I}_1+\mathbb{I}_2+\mathbb{I}_3+\mathbb{I}_4$ with
\begin{eqnarray}
\mathbb{I}_1{\equiv}-Q_{1111}-Q_{1112}-Q_{1211}-Q_{1212},\nonumber \\
\mathbb{I}_2{\equiv}-Q_{2211}-Q_{2212}-Q_{2111}-Q_{2112},\nonumber \\
\mathbb{I}_3{\equiv}-Q_{1221}-Q_{1222}-Q_{1121}-Q_{1122},\nonumber \\
\mathbb{I}_4{\equiv}-Q_{2121}-Q_{2122}-Q_{2221}-Q_{2222}.
\end{eqnarray}
For four qudits, the system may consist of three-qudit and one-qudit subsystems, or of two two-qudit subsystems when we study partial entanglement.
For the former case, define
$r_{ijkl}{\equiv}\xi_{ijk}+\zeta_l$ and write $\mathbb{I}_1$ as
\begin{eqnarray}
\mathbb{I}_1&=&-g_1(r_{1111}+3)-g_2(r_{1112}+3)\nonumber\\
&&-g_2(r_{1211}+3)-g_1(r_{1212})\nonumber\\
&=&-g_1(\xi_{111}+\zeta_1+3)-g_2(\xi_{111}+\zeta_2+3)\nonumber\\
&&-g_2(\xi_{121}+\zeta_1+3)-g_1(\xi_{121}+\zeta_2).
\end{eqnarray}
If we set $\xi_{111}=\alpha_1$, $\xi_{121}=\alpha_2$,
$\zeta_{1}=\beta_1$, $\zeta_{2}=\beta_2$, we find that
$\mathbb{I}_1$ is equivalent to $I^{2}$ and so $\mathbb{I}_1\leq2$;
similarly if we set $\xi_{221}=\alpha_1+3$, $\xi_{211}=\alpha_2$,
$\zeta_{1}=\beta_1$, $\zeta_{2}=\beta_2$, then $\mathbb{I}_2\leq2$;
if we set $\xi_{122}=\alpha_1+3$, $\xi_{112}=\alpha_2$,
$\zeta_{1}=\beta_1$, $\zeta_{2}=\beta_2$, then $\mathbb{I}_3\leq2$;
if we set $\xi_{212}=\alpha_1+3$, $\xi_{222}=\alpha_2+3$,
$\zeta_{1}=\beta_1$, $\zeta_{2}=\beta_2$, then $\mathbb{I}_4\leq2$,
and thus we find
$I^{[4]}=\mathbb{I}_1+\mathbb{I}_2+\mathbb{I}_3+\mathbb{I}_4\leq8$.
For the latter, we define $r_{ijkl}{\equiv}\xi_{ij}+\zeta_{kl}$.
If we set $\xi_{11}=\alpha_1$, $\xi_{12}=\alpha_2$,
$\zeta_{11}=\beta_1$, $\zeta_{12}=\beta_2$, then
$\mathbb{I}_1\leq2$; if we set $\xi_{22}=\alpha_1+3$,
$\xi_{21}=\alpha_2$, $\zeta_{11}=\beta_1$, $\zeta_{12}=\beta_2$,
then $\mathbb{I}_2\leq2$; if we set $\xi_{12}=\alpha_1+3$,
$\xi_{11}=\alpha_2$, $\zeta_{21}=\beta_1$, $\zeta_{22}=\beta_2$,
then $\mathbb{I}_3\leq2$; if we set $\xi_{21}=\alpha_1+3$,
$\xi_{22}=\alpha_2+3$, $\zeta_{21}=\beta_1$, $\zeta_{22}=\beta_2$,
then $\mathbb{I}_4\leq2$. Thus
$I^{[4]}=\mathbb{I}_1+\mathbb{I}_2+\mathbb{I}_3+\mathbb{I}_4\leq8$.

\emph{Arbitrary $N$ qudits.}--- Based on the CGLMP
inequality~(\ref{originalCGLMP}), now we can prove the $N$-qudit
Bell-type inequality~(\ref{N-qudit}) as what we have done in the
three- and four-qudit cases. For further convenience and without
losing the generalization, we abbreviate the correlation function
$-Q_\mathcal{I}$ by $(k)$, here $k=t(\mathcal {I})$. For examples,
$(0)$ means that $i_1=\cdots=i_N=1$ in $Q_{i_1i_2\cdots i_N}$, or
$(0)=-Q_{11\cdots 1}$; $(1)$ means that one of the index
$\{i_1,i_2,\cdots,i_N\}$ is $2$ and the others are $1$, or
$(1)=-Q_{21\cdots 1}/-Q_{12\cdots 1}/\cdots/-Q_{11\cdots 2}$; and
$(k)$ represents the correlation function $-Q_\mathcal{I}$ in which
there are $k$ $``2"$ and $(N-k)$ $``1"$ in the index $\mathcal{I}$.
It is easy to see that the number of $(0)$ is $C_N^0$, that of $(1)$
is $C_N^1$, and so on. In this language, the four-qudit Bell-type
inequality can be expressed as $I^4=\sum_{j=1}^4\mathbb{I}_j\leq 8$
with
\begin{eqnarray}
&&\mathbb{I}_1{\equiv}(0)+(1)+(1)+(2),\nonumber \\
&&\mathbb{I}_2{\equiv}(1)+(2)+(2)+(3),\nonumber\\
&&\mathbb{I}_3{\equiv}(1)+(2)+(2)+(3),\nonumber\\
&&\mathbb{I}_4{\equiv}(2)+(3)+(3)+(4).
\end{eqnarray}
For $N=4$, the $2^4$ correlation functions are divided into four
subgroups with each subgroup possessing the feature of
\begin{eqnarray}
&& \mathcal {G}(k)\equiv(k)+(k+1)+(k+1)+(k+2)
\end{eqnarray}
 as shown above. One more thing
worth to note is that $\mathbb{I}_j$ are grouped according to the
index string $i_1i_2i_3i_4$. Take $\mathbb{I}_1$ as an example,
$\mathbb{I}_1=-(Q_{111}+Q_{112})\otimes (Q_1+Q_2)$ for the case that
the four-qudit system consists of three-qudit and one-qudit
subsystems; and $\mathbb{I}_1=-(Q_{11}+Q_{12})\otimes
(Q_{11}+Q_{12})$ for the case that the system consists of two
two-qudit subsystems. One can find $t(111)+1=t(112)$ and
$t(1)+1=t(2)$ for the former case, and $t(11)+1=t(12)$ for the
latter case. Similar results can be obtained for other
$\mathbb{I}_j$.


Now let us look at $N$ qudits. First we show that the $N$-qudit
Bell-type inequality can be rearranged into a grouping with each
subgroup of the form $\mathcal {G}(k)$, that is
\begin{eqnarray}
I^N=\sum_{k=0}^{N-2}T(k)\times\mathcal {G}(k),
\end{eqnarray}
here $T(k)$ indicates the times that the element $\mathcal {G}(k)$
appears. We obtain iterative equations for the coefficients $T(k)$
\begin{eqnarray}
&&T(0)=C_N^0=1,\nonumber \\
&&T(1)=C_N^1-2T(0)=C_N^1-2C_N^0,\nonumber \\
&&T(2)=C_N^2-2T(1)-T(0)=C_N^2-2C_N^1+3 C_N^0,\nonumber\\
&&\vdots\nonumber \\
&&T(k)=\sum_{i=0}^{k}(-1)^{k-i}(k+1-i)C_N^i.
\end{eqnarray}
The summation of $T(k)$ yields $\sum_{k=0}^{N-2}T(k)=2^{N-2}$.
Since there are four  terms in each $T(k)$, so the total number of
terms $4\sum_{k=0}^{N-2}T(k)=2^N$ is exactly the number of terms in
a $N$-particle inequality for two settings. Therefore such a
rearrangement always exists in our inequality.

Secondly, consider a $N$-qudit system consisting of two subsystems
of $m$ qudits and $N-m$ qudits. According to our two-setting
scenario, in subsystem $m$ we have a set of $2^{m}$ index strings
$i_1i_2\cdots{i_m}$ and in subsystem $N-m$ we have a set of
$2^{N-m}$ index strings $i_{m+1}\cdots{i_N}$; by connection of index
strings of two subsystems we have totally $2^{m}\cdot2^{N-m}=2^N$
index strings $i_1i_2\cdots{i_m}i_{m+1}\cdots{i_N}$ indicating the
kind of measurement for each qudit in the whole system. This implies
the correlation function $Q_{i_1\cdots i_mi_{m+1}\cdots
i_N}=Q_{i_1\cdots i_m}\otimes Q_{i_{m+1}\cdots i_N}$. Let the four
correlation functions in each subgroup be of the form
\begin{eqnarray}
&&-Q_{i_1\cdots i_mi_{m+1}\cdots i_N}-Q_{i_1\cdots i_mi'_{m+1}\cdots i'_N}\nonumber\\
&&-Q_{i'_1\cdots i'_mi_{m+1}\cdots i_N}-Q_{i'_1\cdots
i'_mi'_{m+1}\cdots i'_N}.\label{4cf}
\end{eqnarray}
As stated above, the rule for grouping correlation functions is
based on the index strings. The general rule is
\begin{eqnarray}
&&t(i_1i_2\cdots i_m)+1=t(i'_1i'_2\cdots i'_m),\nonumber\\
&&t(i_{m+1}\cdots i_N)+1=t(i'_{m+1}\cdots i'_N).
\end{eqnarray}
Use the property
$t(i_1\cdots{i_m}i_{m+1}\cdots{i_N})=t(i_1\cdots{i_m})+t(i_{m+1}\cdots{i_N})$,
and let $k=t(i_1\cdots{i_m}i_{m+1}\cdots{i_N})$, we find
$k+1=t(i_1\cdots{i_m}i'_{m+1}\cdots{i'_N})=t(i'_1\cdots{i'_m}i_{m+1}\cdots{i_N})$,
and $k+2=t(i'_1\cdots{i'_m}i'_{m+1}\cdots{i'_N})$, we then exactly
have the simple form of $\mathcal {G}(k)$ from the expression
(\ref{4cf}).


Next we show that $\mathcal {G}(k)\leq 2$ by considering two cases
of $k={\rm even}$ and $k={\rm odd}$.

Case a: $k$ is even, we obtain
\begin{eqnarray}
&&\mathcal {G}(k)=\nonumber\\
&&g_1(r_{i_1\cdots i_mi_{m+1}\cdots i_N}+3-3[k/2]) \nonumber \\
&&+g_2(r_{i_1\cdots i_mi'_{m+1}\cdots i'_N}+3-3[(k+1)/2]) \nonumber \\
&&+g_2(r_{i'_1\cdots i'_mi_{m+1}\cdots i_N}+3-3[(k+1)/2]) \nonumber \\
&&+g_1(r_{i'_1\cdots i'_mi'_{m+1}\cdots i'_N}+3-3[(k+2)/2]).
\end{eqnarray}
As the $N$-qudit system consists of two subsystems of $m$ qudits and
$N-m$ qudits, we have
$r_{i_1i_2\cdots{i_N}}=\xi_{i_1\cdots{i_m}}+\zeta_{i_{m+1}\cdots{i_{N}}}$.
If we set $\xi_{i_1\cdots{i_m}}=\alpha_1+3a$,
$\xi_{i'_1\cdots{i'_m}}=\alpha_2+3a$,
$\zeta_{i_{m+1}\cdots{i_{N}}}=\beta_1$,
$\zeta_{i'_{m+1}\cdots{i'_{N}}}=\beta_2$, and $a=k/2$, then this
group of correlation functions is equivalent to the CGLMP
inequality~(\ref{originalCGLMP}) and thus its upper bound is 2.

Case b: $k$ is odd, we have
\begin{eqnarray}
&&\mathcal {G}(k)=\nonumber\\
&&g_1(\xi_{i'_1\cdots{i'_m}}+\zeta_{i_{m+1}\cdots{i_{N}}}+3-3[(k+1)/2]) \nonumber \\
&&+g_2(\xi_{i'_1\cdots{i'_m}}+\zeta_{i'_{m+1}\cdots{i'_{N}}}+3-3[(k+2)/2]) \nonumber \\
&&+g_2(\xi_{i_1\cdots{i_m}}+\zeta_{i_{m+1}\cdots{i_{N}}}+3-3[k/2]) \nonumber \\
&&+g_1(\xi_{i_1\cdots{i_m}}+\zeta_{i'_{m+1}\cdots{i'_{N}}}+3-3[(k+1)/2]).
\end{eqnarray}
If we set $\xi_{i_1\cdots{i_m}}=\alpha_2+3(b-1)$,
$\xi_{i'_1\cdots{i'_m}}=\alpha_1+3b$,
$\zeta_{i_{m+1}\cdots{i_{N}}}=\beta_1$,
$\zeta_{i'_{m+1}\cdots{i'_{N}}}=\beta_2$, and $b=(k+1)/2$, then
$\mathcal {G}(k)$ is equivalent to the CGLMP
inequality~(\ref{originalCGLMP}) and thus its upper bound is 2.




Based on the above analysis, such a group of correlation functions
$\mathcal {G}(k)$ is always less than 2. There are totally $2^{N-2}$
subgroups in $I^{N}$, therefore $I^{N}\le 2^{N-1}$. This ends the
proof.

\section{quantum violation of the $N$-qudit Bell-type inequality}
\label{s3} We now turn to study quantum violations of the inequality
(\ref{N-qudit}) for the GHZ states 
\begin{eqnarray}
|\psi \rangle^{N}_{\rm GHZ}=\frac{1}{\sqrt{d}}\sum_{j=0}^{d-1}|jj\cdots
j\rangle,\label{GHZ}
\end{eqnarray}
which are fully entangled states of $N$-qudit. Quantum mechanical
joint probability is calculated by
\begin{eqnarray}
&&P^{{\rm QM}}(x_{i_1},x_{i_2},\cdot\cdot\cdot,x_{i_N})
=\texttt{Tr}[\rho^{N}(U_{i_1}\otimes
U_{i_2}\otimes\cdots\otimes U_{i_N})\nonumber\\
&&\times
(\Pi_{i_1}\otimes\Pi_{i_2}\otimes\cdots\otimes\Pi_{i_N})(U^{\dagger}_{i_1}\otimes
U^{\dagger}_{i_2}\otimes\cdots\otimes U^{\dagger}_{i_N})],
\end{eqnarray}
where $\Pi_{i_n}=|x_{i_n}\rangle\langle x_{i_n}|$ and $U_{i_n}
\;(n=1,2,...,N)$ are projectors and unitary transformation operators
for corresponding qudits. As for $U_{i_n}$, it is sufficient to
consider the unbiased symmetric multi-port
beamsplitters~\cite{1997M. Zukowski-Z.Zeilinger} when we study the
GHZ states. A photon entering at any input ports of a symmetric
$d$-port beam splitter has an equal chance as $1/d$ of exiting at
any output ports. The action of the multiport beam splitter can be
described by a unitary transformation $\mathcal {T}$ with elements
$\mathcal {T}_{kl}=\frac{1}{\sqrt{d}}\omega^{kl}$, where
$\omega=\exp(\frac{i2\pi}{d})$. In front of the $i$-th input port,
there is a phase shifter to adjust the phase of the incoming photon
by $\phi_i$. The phase shifts can be denoted as a $d$-dimensional
vector $\vec{\phi}=(\phi_0,\phi_1,\cdots,\phi_{d-1})$. The symmetric
$d$-port beam splitter together with the $d$ phase shifters perform
the unitary transformation $U_{i_n}(\vec{\phi})$ with elements
$U_{kl}(\vec{\phi})=\frac{\omega^{kl}\exp(i\phi_{l})}{\sqrt{d}}$.
Specifically, take the first qudit as an example, the phase angles
are $\vec{\phi}_{i_1=1}=(\phi_{10},\phi_{11},\cdots,\phi_{1(d-1)})$
and $\vec{\phi}_{i_1=2}=(\phi_{20},\phi_{21},\cdots,\phi_{2(d-1)})$
due to the two-setting scenario.

Numerical calculations show that
maximal violation of the Bell-type inequality for the GHZ states can
be found with the following optimal angle settings
\begin{eqnarray}
\vec{\phi}_{i_n=1}&=&\vec{\phi}_{i_1=1}\nonumber\\
&=&(0,\frac{m_1\pi}{2d},2\frac{m_1\pi}{2d},\cdots,(d-1)\frac{m_1\pi}{2d}),\\
\vec{\phi}_{i_n=2}&=&\vec{\phi}_{i_1=2}\nonumber\\
&=&(0,\frac{m_2\pi}{2d},2\frac{m_2\pi}{2d},\cdots,(d-1)\frac{m_2\pi}{2d}),
\end{eqnarray}
where $m_1=15/N$, $m_2=m_1-6$. 
For $N=2$, our result correctly recovers that of
the CGLMP inequality \cite{2002CGLMPBI}, namely
\begin{eqnarray}
[I^{2}]^{\rm max}=4d
\sum_{k=0}^{[d/2]-1}(1-\frac{2k}{d-1})(q_k-q_{-(k+1)}),
\end{eqnarray}
where $q_c=\frac{1}{2 d^3 \sin^2[\pi(c+1/4)/d]}$. The maximal
violations increase with dimension $d$, for examples,
$[I^{2}_{d=2}]^{\rm max}=2 \sqrt{2}\simeq 2.828$ and
$[I^{2}_{d=3}]^{\rm max}=(12+8\sqrt{3})/9\simeq 2.873$. For
arbitrary $N$-qudit, the maximal violation is
\begin{eqnarray}
[I^{N}]^{\rm max}=2^{N-2} \times [I^{2}]^{\rm max}.
\label{violation}
\end{eqnarray}
One can show how sensitive the inequality (\ref{N-qudit}) is by
considering the factor $\mathcal {R}$ defined by maximal violation
of the inequality over upper bound for true $N$-body
entanglement \cite{2005M. Zukowski}, i.e., 
\begin{eqnarray}
\mathcal {R}=[I^{N}]^{\rm max}/2^{N-1}. \label{factor}
\end{eqnarray}
It is easy to have $\mathcal {R}=\frac{[I^{2}]^{\rm max}}{2}$ for
two qudits, and this confirms that inequality (\ref{N-qudit}) is an
equivalent form of the CGLMP inequality when $N=2$. For $d=2$, the
Bell-type inequality (\ref{N-qudit}) is an equivalent version of the
Svetlichny inequality for $N$ qubits, and accordingly $\mathcal
{R}=\sqrt{2}$. Our result is in accordance with Refs.
\cite{2002Seevinck,2002Collins} (see Eq. (14) in
\cite{2002Collins}).

So far we have discussed the violations of the pure GHZ states. If
white noise is added, the pure state turns to a mixed state as
\begin{eqnarray}
\rho^N(V)= V \rho_{\rm GHZ}+ (1-V) \rho_{\rm noise}, \label{noise}
\end{eqnarray}
where $\rho_{\rm noise}=\frac{\textbf{1}}{d^N}$, $\textbf{1}$ is the
unit operator, $V$ is the so-called visibility, and $0 \le V \le 1$.
We find that the mixed state violates inequality (\ref{N-qudit}) if
$V
> V_{\rm cr}$, where $ V_{\rm cr}= \frac{1}{\mathcal {R}}$ is the
critical value of visibility. The critical values decrease when
dimension $d$ goes up, for examples,
$V_{\rm cr}=0.707$ for $d=2$ and $V_{\rm cr}=0.696$ for $d=3$. The HLNHV description of the state 
is not allowed if $V > V_{\rm cr}$.
The less the value of $ V_{\rm cr}$ is, the more noise tolerant the Bell inequality is.
Moreover, for any partially entangled states
\begin{eqnarray}
\rho^N=\rho^m\otimes\rho^{N-m},
\end{eqnarray}
the quantum joint probability are factorizable, i.e.,
\begin{eqnarray}
&&P^{{\rm QM}}(x_{i_1},x_{i_2},\cdots,x_{i_N})=P_1^{{\rm
QM}}(x_{i_1},x_{i_2,\cdots,x_{i_m}}) \nonumber\\
&&\;\;\;\;\;\;\;\;\;\;\;\;\;\;\;\;\;\;\;\;\;\;\;\times P_2^{{\rm
QM}}(x_{i_{m+1}},x_{i_{m+2}},\cdots,x_{i_N}).
\end{eqnarray}
Consequently, our inequality holds for any partially entangled
states or any convex mixture of them. Any violation of our
inequality is a sufficient condition to confirm full $N$-particle
entanglement and rule out the HLNHV description. Let us point out
that the $N$-qubit Svetlichny inequality can also be used to test
the invalidity of the HLNHV description for $N$ qudits by dividing
the $d$ outcomes into two sets, provided that $V>V_{\rm cr}^{\rm s}$
with $V_{\rm cr}^{\rm s}$ being the critical visibility for the
Svetlichny inequality. It is easy to see that $V_{\rm cr}^{\rm s}=
\frac{1}{\sqrt{2}}$ does not depend on dimension $d$, and $V_{\rm
cr}<V_{\rm cr}^{\rm s}$ when $d\geq 3$. Thus our Bell-type
inequality is more noise resistant than the Svetlichny one for $N$
qudits when $d\geq 3$.

\section{Conclusion}
\label{s4}
Based on the assumption of
partial separability we have derived a set of Bell-type inequalities
for arbitrarily high-dimensional systems. Partially entangled states
would not violate the inequalities, thus upon violation, the
Bell-type inequalities are sufficient conditions to detect the full
$N$-qudit entanglement and rule out the HLNHV description. It is observed that the
Bell-type inequalities for multi-qudit ($d \ge 3$) violate the
hybrid local-nonlocal realism more strongly than the Svetlichny ones
for qubits, and the quantum violations increase with dimension $d$.
Furthermore, how to generalize the Bell-type inequality to the
multi-setting one remains a significant topic to be investigated further.

We thank Profs. N. Gisin, S. Popescu and N. Brunner for their
helpful remarks and discussions. J.L.C. is supported by NSF of China
(Grant No. 10975075) and the Fundamental Research Funds for the
Central Universities. This work is also partly supported by National Research Foundation and
Ministry of Education, Singapore (Grant No. WBS: R-710-000-008-271).

\emph{Note added.}--- Ref. \cite{gisin2010} also presents a similar
inequality to our inequality (\ref{N-qudit}).



\begin{thebibliography}{99}


\bibitem{Benne1} C. H. Bennett, G. Brassard, C. Cr{\' e}peau, R. Jozsa, A. Peres, and W. K. Wootters, Phys. Rev. Lett. \textbf{70}, 1895 (1993).


\bibitem{Gisin1} N. Gisin, G. Ribordy, W. Tittel, H. Zbinden, Rev. Mod. Phys. \textbf{74}, 145 (2002).

\bibitem{1987Svetlichny} G. Svetlichny, Phys. Rev. D \textbf{35}, 3066 (1987).


\bibitem{2002Seevinck} M. Seevinck and G. Svetlichny, Phys. Rev. Lett. \textbf{89}, 060401 (2002).


\bibitem{2002Collins} D. Collins, N. Gisin, S. Popescu, D. Roberts, and V. Scarani, Phys. Rev. Lett. \textbf{88}, 170405 (2002).


\bibitem{2009Ghose} S. Ghose, N. Sinclair, S. Debnath, P. Rungta, and R. Stock, Phys. Rev. Lett. \textbf{102}, 250404 (2009).

\bibitem{Rpreprint} O. G\"{u}hne and M. Seevinck, New J. Phys. \textbf{12}, 053002 (2010).



\bibitem{2005M. Zukowski} W. Laskowski and M. \.{Z}ukowski, Phys. Rev. A \textbf{72}, 062112 (2005).


\bibitem{Rauschenbeutel} A. Rauschenbeutel, G. Nogues, S. Osnaghi, P. Bertet, M. Brune, J. Raimond, and S. Haroche, Science \textbf{288}, 2024 (2000).


\bibitem{JWPan} J.-W. Pan, D. Bouwmeester, M. Daniell, H. Weinfurter, and A. Zeilinger, Nature (London) 403, 515 (2000).

\bibitem{Bell} J. S. Bell, Physics (Long Island City, N.Y.) \textbf{1}, 195 (1964).


\bibitem{MABK} N. D. Mermin, Phys. Rev. Lett. \textbf{65}, 1838 (1990); M.
Ardehali, Phys. Rev. A \textbf{46}, 5375 (1992); A. V. Belinskii and
D. N. Klyshko, Phys. Usp. \textbf{36}, 653 (1993).

\bibitem{Werner1} R. F. Werner and M. M. Wolf, Phys. Rev. A \textbf{64}, 032112 (2001).


\bibitem{Zukowski1} M. \.{Z}ukowski and \v{C}. Brukner, Phys. Rev. Lett. \textbf{88}, 210401 (2002).


\bibitem{2004Mitchell-three-particle-nonlocality} P. Mitchell, S. Popescu, and D. Roberts, Phys. Rev. A \textbf{70}, 060101(R) (2004).

\bibitem{1998Percival}
I. Percival, Phys. Lett. A \textbf{244}, 495 (1998).

\bibitem{1982Fine}
A. Fine, Phys. Rev. Lett. \textbf{48}, 291 (1982).


\bibitem{2002CGLMPBI} D. Collins, N. Gisin, N. Linden, S. Massar, and S. Popescu, Phys. Rev. Lett. \textbf{88}, 040404 (2002).


\bibitem{1997M. Zukowski-Z.Zeilinger} M. \.{Z}ukowski, A. Zeilinger, and M. A. Horne, Phys. Rev. A \textbf{55}, 2564 (1997).


\bibitem{gisin2010} J.-D. Bancal, N. Brunner, N. Gisin, and Y.-C. Liang, Phys. Rev. Lett.
{\bf 106}, 020405(2011). arXiv:1011.0089.


\end{thebibliography}
\end{document}